\documentclass[prd,nofootinbib,amsfonts]{revtex4}
\usepackage{graphicx,bm,amsmath,color}
\usepackage[latin1]{inputenc}

\newcommand{\be}{\begin{equation}}
\newcommand{\ee}{\end{equation}}
\newcommand{\xbf}{\bm{x}}
\newcommand{\pbf}{\bm{p}}

\newcommand{\pibf}{\bm{\pi}}

\begin{document}

\title{Asymptotic approach to Special Relativity compatible with a
  relativistic principle}
\author{J.M. Carmona}
\email{jcarmona, cortes, dmazon@unizar.es}
\affiliation{Departamento de F\'{\i}sica Te\'orica,
Universidad de Zaragoza, Zaragoza 50009, Spain}
\author{J.L. Cort\'es}
\email{jcarmona, cortes, dmazon@unizar.es}
\affiliation{Departamento de F\'{\i}sica Te\'orica,
Universidad de Zaragoza, Zaragoza 50009, Spain}
\author{D. Maz\'on}
\email{jcarmona, cortes, dmazon@unizar.es}
\affiliation{Departamento de F\'{\i}sica Te\'orica,
Universidad de Zaragoza, Zaragoza 50009, Spain}
\begin{abstract}

We propose a general framework to describe Planckian
deviations from Special Relativity (SR) compatible with a relativistic
principle. They are introduced as the
leading corrections in an asymptotic approach to SR going beyond the
energy power expansion of effective field theories. We discuss the
conditions in which these
Planckian effects might be experimentally observable
in the near future, together with the non-trivial limits of
applicability of this asymptotic approach that such a situation would
produce, both at the very high (ultraviolet)
and the very low (infrared) energy regimes.

\end{abstract}

\maketitle

\section{Introduction}

Lorentz invariance is a main ingredient in our present day physical
theories. However, Planck-scale departures from Lorentz symmetry are
naturally expected as quantum gravity effects in many explored
scenarios in relation with the structure of a quantum
space-time~\cite{qugra}. It is interesting to realize that going
beyond Special Relativity (SR) does not necessarily imply the
existence of a preferred class of inertial observers. Doubly Special
Relativity (DSR) was originally proposed as a generalization of SR in
which the key idea is to have a universal, observer-independent,
length (mass) scale besides the universal maximum speed present in
SR. In this way one could give to the Planck length, made invariant by
a suitable deformation of the Lorentz transformations between inertial
observers, a physical meaning in connection with the structure of
space-time~\cite{dsr,KG,magueijo}.

DSR has been studied for almost ten years now~\cite{dsrfacts}. It has
been shown that there exist different versions of it~\cite{dsr,dsr2,dsr3},
in which the universal energy scale represents typically a bound of
the energy or/and the momentum. However, the consistency of a full theory
implementing the DSR idea is still an open question, with some doubts
coming from the emergence of apparent paradoxes, such as the so-called
\textit{soccer ball} problem (i.e., the
incompatibility
between the kinematics of macroscopic bodies and the assumption of
a Planckian deformed dispersion relation)~\cite{soccerball,magueijo},
or nonlocal effects
that might affect macroscopic physics~\cite{nonlocality}.

One might wonder whether the DSR idea could be tested. In fact it may
appear surprising that quantum gravity effects may be accesible at all
to present day experiments, which involve energies many orders of
magnitude lower than the Planck scale. However, the realization that
in certain situations there are mechanisms that are able to amplify
these corrections has opened a whole branch in the field, known
generically as quantum gravity phenomenology~\cite{qugraphen}.
It is in principle possible that this phenomenology is able to
distinguish a DSR from a Lorentz breakdown (without a relativity
principle) scenario. For example, by experimentally
establishing the existence of an energy threshold for some particle decay the
DSR framework could be ruled out, since such a threshold would not be
observer independent~\cite{dsrfacts}. Nevertheless, in a general case it
will be difficult to distinguish between both scenarios. For example,
most DSR versions give negligible corrections for processes
typically considered in quantum gravity phenomenology, such as the
analysis of the ultra-high energy cosmic-ray spectrum and the presence
of the GZK cutoff~\cite{GZK}, but there is always the possibility to cook some
ad hoc DSR model making the corrections relevant~\cite{dsr3}.
What is more, at least in the next
future, it seems likely that this kind of phenomenology will be
sensitive at most to the leading order of DSR or Lorentz breakdown
corrections, so that one should not insist on pursuing an ``all-order
DSR phenomenology''~\cite{dsrfacts}.

Taking into account these considerations, and with the same motivation
that inspired the DSR idea, we try in this work to present a
systematic, very general and coherent framework of asymptotic
Planck-scale corrections to Special Relativity in a relativistic
invariant way. These corrections, parametrized by the Planck scale,
are a first-order approximation to whatever scenario lies beyond SR,
so that their validity will be limited to an energy domain which will
depend on the details of the implementation of the departures from SR
kinematics. Since this domain will in any case lie far from the Planck
energy, we will not impose any restriction of the kind of the DSR
postulate of having the Planck mass as a second invariant besides the
speed of light at the level of this asymptotic approach.
This fact will also eliminate the need to address
any soccer ball problem at this level since it would imply to go
beyond the limits of validity of the approximation we are
considering. We will refer to this framework as Asymptotic Special
Relativity (ASR). It will contain of course the leading order of every
DSR theory, but ASR maybe relevant even if DSR (as an all-order
theory) is not realized in Nature.

Implicit in all the discussion presented here is the
assumption that the framework of effective field theories is not
sufficiently general to incorporate all the
low-energy/semiclassical implications of the
quantum structure of space-time. If this were not the case then
one already has
an extension of the Standard Model in the effective field theory
(EFT)
framework~\cite{esm} incorporating physics beyond SR.
A consequence of this assumption is that one is limited to
kinematic considerations. We are still far from identifying a
consistent dynamical theory including the main consequences of quantum
gravity which allows us to understand how SR emerges as an approximate
or exact symmetry, but we think there are enough reasons to
explore kinematic situations which go beyond the EFT paradigm.
In fact, EFT has shown to be a powerful tool in describing a large variety of
phenomena,
from the Fermi theory of weak interactions in particle physics to the quantum
Hall effect in condensed matter, just to mention two important examples, but
this does not mean that every physical effect can be explained in its framework.
In condensed matter physics the description of many-body
systems in terms of a few collective degrees of freedom (the quasi-particles) is
always approximate and, in many cases, the construction of a local EFT
is impossible due to some exchange with the fundamental degrees of freedom
(the atoms)~\cite{drawbacksEFT}. This would be relevant in the low-energy/semiclassical
limit of quantum gravity
if gravitons were emergent degrees of freedom~\cite{analogues}. In other
approaches to the
quantum gravity problem there are also hints pointing out that
their low-energy/semiclassical limit is not completely contained in a local EFT.
In quantum field theories with canonical noncommutative
space-time coordinates there exists a mixing between ultraviolet and infrared modes
that makes difficult its compatibility with the Wilsonian EFT
framework~\cite{UV/IR mixing}.
In approaches independent of background metrics, such as Loop
Quantum Gravity, where a different quantization (i.e., not
unitarily equivalent
to the Fock's quantization) is carried out and radically
new ingredients arise, it is not clear whether
the low-energy/semiclassical limit of the theory is completely
contained in a
conventional local EFT~\cite{BI}.

In fact, the approach introduced in this paper under the name of ASR is
not new. The possibility of considering DSR as a leading order
Planckian effect independently of an extension to all orders and
energies was already present in the first works of
DSR~\cite{dsr,magueijo}. What we present in this paper is a general and
systematic discussion of the kind of corrections consistent with the
existence of a relativity principle and their phenomenological
consequences, something we think was missing in previous
works. Besides, we introduce some notions such as an infrared limit
(additional to the ultraviolet one) in the domain of validity of the
departures from SR and an intrinsic non-separability
in the energy-momentum conservation law.

In the next Section we present a generalization of SR kinematics for
one and two particle systems. In Section III we explore the
observability of the departures from SR with a few examples. In
Section IV we summarize our results and proposals for future work.

\section{General kinematic framework}

\subsection{One-particle system}

We start with the kinematic description of a one particle system. It is
based on the assumption of translational invariance, which allows to
assign energy-momentum variables $E$, ${\bf P}$ to each particle, and
rotational invariance implemented as in SR (most attempts to go beyond
SR concentrate on this possibility,\footnote{There are several
arguments to consider on a different footing deviations from SR at the
level of rotation/boost transformations. One is the compact/noncompact
nature of the transformations which prevents the possibility to test
the invariance under all boost transformations. Another one is the
observation that the presence of a new
energy scale in the modified dispersion relation requires
deforming the boost transformations but not necessarily the rotations.} although it
is possible to go beyond this case). The main ingredient defining the
kinematic framework is its asymptotic approach to SR kinematics
parametrized by just one single additional scale ($M_{P}$, the scale
of transition from the classical to the quantum description of
space-time). Under these conditions the most general
expression for the
asymptotic dispersion relation of a particle takes the form
\be
E^2 - P^2 - \frac{E^{2+\alpha}}{M_P^\alpha}
\Delta(P/E) \,=\, \mu^2
\label{adr}
\ee
with $\mu$ the particle mass, which in the region where
Planckian
corrections are completely negligible coincides with the SR
definition, $\mu^2=E^2-P^2$, an exponent $\alpha>0$ characterizing the
leading approach to SR when
$E/M_P\ll 1$, and an arbitrary universal\footnote{As long as we are interested
in an implementation of the relativity principle, we will not consider
dispersion relations that depend on particle properties, apart from
the mass.}, i.e. the same for every particle,
function $\Delta$ of the dimensionless
combination of rotational invariant variables $P/E$. The Planckian
modified dispersion relations which appear in many phenomenological
studies of departures from SR are included as
particular cases (with $\alpha=1$ or $\alpha=2$, and specifics forms of $\Delta$,
such us a constant value) of the asymptotic dispersion relation~(\ref{adr}).

In order to make this kinematic description compatible with a relativity principle
we need two things. First, identify a set of inertial frames for which
some deformed Lorentz transformations
specify the relation between energy-momentum variables in them in such
a way that the dispersion relation (\ref{adr})
is kept invariant. Second, modify accordingly the usual conservation
laws so that they become covariant
under change of inertial frames. The last point requires going beyond
the one particle system
and will be studied in the next Subsection. The simplest way to satisfy
the first requisite is to introduce
a set of auxiliary asymptotic variables
\be
\epsilon = E \left(1 - \frac{E^\alpha}{2 M_P^\alpha}
g(P/E)\right) {\hskip 1cm}
\pibf = {\bf P} \left(1 - \frac{E^\alpha}{2 M_P^\alpha}
h(P/E)\right)
\label{aav}
\ee
where the two functions $g$, $h$ defining the auxiliary variables are
chosen such that
\be
g(P/E) - \frac{P^2}{E^2} \, h(P/E) \,=\, \Delta(P/E)\,.
\ee
Apart from this, there is no other restriction on these two functions but the trivial
request that there must be some range of $P/E$ values where they are bounded.
The different choices for $g$ and $h$ will determine the domain of
validity of the asymptotic approach to SR, to be discussed in the next
paragraphs. One can then write the asymptotic dispersion relation
(\ref{adr}) in the simple form
\be
\epsilon^2 - \pi^2 \,=\, \mu^2\,,
\label{disprelaux}
\ee
and the Lorentz transformation of energy-momentum variables in
different inertial frames is obtained by translating the usual linear
four-vector Lorentz transformations of the auxiliary asymptotic
variables $\epsilon$, $\pibf$ to the energy-momentum of the
particle through the relations
\be
E = \epsilon \left(1 + \frac{\epsilon^\alpha}{2 M_P^\alpha}
g(\pi/\epsilon)\right) {\hskip 1cm}
{\bf P} = \pibf \left(1 + \frac{\epsilon^\alpha}{2 M_P^\alpha}
h(\pi/\epsilon)\right)\,.
\label{epav}
\ee
The asymptotic auxiliary variables are the analog of the `Judes-Visser
auxiliary variables'~\cite{jv} introduced in DSR. Is is important
to remark that given a
dispersion relation there are several possible choices of functions
$g$ and $h$ compatible with the implementation of the relativity principle,
each one corresponding to a transformation between inertial reference frames.

Then one has that the kinematic description of a one particle system
in a generalization of SR is given in terms of an exponent $\alpha$
parametrizing the asymptotic approximation to SR and two functions
$g$, $h$ of one variable specifying the transformation of
energy-momentum variables among different frames and the asymptotic
dispersion relation.

\subsubsection*{Domain of validity of ASR in the one-particle system}

The asymptotic approach to SR is governed by $(E/M_P)^{\alpha} g(P/E)$
and $(E/M_P)^{\alpha} h(P/E)$ which control the departure of the
auxiliary variables (in which Lorentz transformations act
linearly)  from the energy momentum variables (in which
Lorentz boosts act non-linearly). Since all our observations are
constrained to energies far below the Planck mass $M_P$ then we will
not have observable signals of the departures from SR if the functions
$g(P/E)$, $h(P/E)$ are of order one. This is, in fact, the general situation that one
has in DSR. On the contrary, in Lorentz breaking theories, which do not preserve the
relativity principle, large observable effects are produced by very small
changes in the dispersion relations. The reason of this different behaviour is due to
inevitable cancellations owing to the consistency between the
dispersion relation and the energy
and momentum conservation laws with the relativity principle. In order to compensate
them we have to consider
observations corresponding to some kinematic limits where the
functions $g(P/E)$, $h(P/E)$ are much larger than one and the
deviations from SR are amplified. This will affect the naive limit of applicability
of ASR, $E\ll M_P$, restricting it to the domain where the difference between
auxiliary and energy momentum variables is small (since ASR gives only asymptotic
corrections to SR).
Let us see how these amplifications might
arise in two kinematic limits, the ultra-relativistic limit $P/E\to 1$ and the
non-relativistic limit $P/E\to 0$.

In the ultra-relativistic limit one can consider a situation where
$g(x)$ (or $h(x)$, or both) has the following behavior:
\be
g(x)\to (1-x)^{-\gamma},\ \gamma>0, \quad \text{when } x\to 1.
\label{URlimit}
\ee
One has in this case
\be
g(P/E)\sim \frac{(1+P/E)^\gamma}{\left(1-P^2/E^2\right)^\gamma}\sim
\left(1+\frac{P}{E}\right)^\gamma \left(\frac{E}{\mu}\right)^{2\gamma}
\label{ginURlimit}
\ee
since $\mu^2=E^2-P^2$ at leading order in the asymptotic correction, so that one gets
an amplification factor $(E/\mu)^{2\gamma}$ for the Planckian corrections which can
make them appreciable at subPlanckian energy scales. At the same time
the domain of validity of the asymptotic description of the kinematics
is now restricted to $E\ll M_P (\mu/M_P)^{1/(1+\alpha/2\gamma)}$. In
fact the DSR scenario~\cite{dsr3} with sizeable threshold
anomalies is just a particular realization of this amplification
mechanism.

In a similar way one can consider the possibility of an amplification
of the Planckian kinematic corrections in the nonrelativistic limit.
In this case a limiting behavior for $g(x)$ (and/or $h(x)$) such as
\be
g(x)\to x^{-\beta},\ \beta>0, \quad \text{when } x\to 0,
\label{NRlimit}
\ee
leads to an amplification factor
$(\mu/P)^\beta$ which can partially compensate the natural suppression
factor $(\mu/M_P)^\alpha$ of Planckian kinematic corrections to the
nonrelativistic limit. We will show in the next Section some examples
along these lines which lead to observable effects in atomic
interferometry. A remarkable property of this scenario is that the
domain of validity of the asymptotic kinematic description is
constrained not only in the UV but also in the IR  by
$(\mu/M_P)^\alpha (\mu/P)^\beta \ll 1$, i.e. $P\gg \mu
(\mu/M_P)^{\alpha/\beta}$. This restriction is
satisfied
by any microscopic experiment up to the present time, a
necessary condition to make
this scenario not obviously phenomenologically inconsistent. On the
other hand an amplification of kinematic Planckian corrections in the
nonrelativistic limit may be a clue to look for sufficiently precise
observations to be able to identify a small departure from SR.

\subsubsection*{Special case: particles of zero mass}

The above discussion is limited to the case where $\mu\neq 0$. For
example, we see that
Eq.~(\ref{ginURlimit}) is problematic if $\mu=0$. In fact, the
generalization of SR to an
asymptotic approximation in terms of the two functions $g$, $h$ of one
variable has to be
changed if we are dealing with a particle for which $\mu=0$. For such
a particle, Eq.~(\ref{disprelaux})
gives $\epsilon=\pi$, so that the quotient $\pi/\epsilon$ is always
equal to one. This implies
that in relations~(\ref{epav}), $g(\pi/\epsilon)$ and
$h(\pi/\epsilon)$ are not functions, but
constants. Let us call them $\tilde{g}$ and $\tilde{h}$, respectively.

The expressions~(\ref{aav}) are then rewritten as
\be
\epsilon = E \left(1 - \frac{E^\alpha}{2 M_P^\alpha}
\tilde{g}\right) {\hskip 1cm}
\pibf = {\bf P} \left(1 - \frac{E^\alpha}{2 M_P^\alpha}
\tilde{h}\right),
\ee
and $\epsilon^2-\pi^2=0$ leads to the dispersion relation
\be
E^2 - P^2 - \frac{E^{2+\alpha}}{M_P^\alpha}
\left(\tilde{g}-\frac{P^2}{E^2}\tilde{h}\right) \,\approx\,
E^2 - P^2 - \frac{E^{2+\alpha}}{M_P^\alpha}
\left(\tilde{g}-\tilde{h}\right)
\,=\, 0\,,
\label{adr0}
\ee
where we have used that in the asymptotic correction $P\approx E$.
That is, the dispersion
relation for a zero mass particle in ASR is like Eq.~(\ref{adr}),
where $\mu=0$ and $\Delta(P/E)$ is a constant.

In this case we do not have at our disposal a function which can
produce amplification effects
in certain kinematic limits, so the domain of validity of ASR for a
zero mass particle will be
given just by $E\ll M_P$. Therefore in particle reactions involving
photons we will not consider
deviations from SR for them. Only at the end of the paper we will come to a possible
phenomenological consequence of the tiny deviations from SR for photons in their
propagation through very large distances.

\subsection{Conservation laws in a two particle system}

To complete the kinematic analysis one needs to identify the
modification of the energy-momentum conservation laws of SR which are
compatible with the generalized relativity principle. The simplest way
to do that is to consider the set of auxiliary asymptotic variables
$\epsilon$, $\pibf$ for each particle. In order to have asymptotic
conservation laws compatible with the relativity principle one has to
identify the most general form to get a set of four variables out of
the auxiliary variables of the different particles in the initial or
final state transforming under Lorentz transformations like the
auxiliary asymptotic variables of a one particle system. In the case
of a two particle system this amounts to get a four-vector out of two
four-vectors, i.e., to consider a linear combination of the two
four-vectors with scalar coefficients
\be
\epsilon_{12} \,=\, \epsilon_1 + \epsilon_2 +
\frac{s^{\alpha/2}}{M_P^{\alpha}}
\left[\epsilon_1 f\left(\frac{\mu_1^2}{s}, \frac{\mu_2^2}{s}
\right) + \epsilon_2 f\left(\frac{\mu_2^2}{s}, \frac{\mu_1^2}{s}
\right) \right]
\label{auxenergyconslaw}
\ee
\be
\pibf_{12} \,=\, \pibf_1 + \pibf_2 + \frac{s^{\alpha/2}}{M_P^{\alpha}}
\left[\pibf_1 f\left(\frac{\mu_1^2}{s}, \frac{\mu_2^2}{s}
\right) + \pibf_2 f\left(\frac{\mu_2^2}{s}, \frac{\mu_1^2}{s}
\right) \right]
\ee
with
\be
s = \mu_1^2 + \mu_2^2 + 2(\epsilon_1\epsilon_2 - \pibf_1 \pibf_2).
\ee
If one does not go beyond a two particle system then the asymptotic
generalized conservation laws can be expressed as an equality between
$\epsilon_{12}$ and $\pibf_{12}$ for the initial and final
states, which are a combination of the auxiliary energy and momentum variables
of the two particles whose general form is determined by a function $f$ of two
variables.
In order to arrive to this result we have also added the
assumptions that there is neither ordering ambiguity,
nor
\textit{initial/final state mixing}. The absence of ordering ambiguity
implies that the
conservation laws do not depend on what particle we choose as the first or the second
particle and the lack of initial/final state mixing means that
the conservation laws can be expressed as a system of equations where
the dependence on initial and final
state variables can be separated.\footnote{More
exotic
possibilities have been considered~\cite{kappa} in the attempts to
realize the DSR idea in the $\kappa$-Minkowski/$\kappa$-Poincar\'e
framework.} The four combinations of variables which appear in the
asymptotic generalized conservation laws when expressed in terms of
the energy-momentum variables of the particles take the form
\be
\epsilon_{12} \,=\, E_1 \left(1 - \frac{E_1^\alpha}{2 M_P^\alpha}
g(P_1/E_1)\right)
+ E_2 \left(1 - \frac{E_2^\alpha}{2 M_P^\alpha}
g(P_2/E_2)\right) +
\frac{s^{\alpha/2}}{M_P^{\alpha}}
\left[E_1 f\left(\frac{\mu_1^2}{s}, \frac{\mu_2^2}{s}
\right) + E_2 f\left(\frac{\mu_2^2}{s}, \frac{\mu_1^2}{s}
\right) \right]
\label{energyconslaw}
\ee
\be
\pibf_{12} \,=\, {\bf P}_1 \left(1 - \frac{E_1^\alpha}{2 M_P^\alpha}
h(P_1/E_1)\right)
+ {\bf P}_2 \left(1 - \frac{E_2^\alpha}{2 M_P^\alpha}
h(P_2/E_2)\right)
+ \frac{s^{\alpha/2}}{M_P^{\alpha}}
\left[{\bf P}_1 f\left(\frac{\mu_1^2}{s}, \frac{\mu_2^2}{s}
\right) + {\bf P}_2 f\left(\frac{\mu_2^2}{s}, \frac{\mu_1^2}{s}
\right) \right],
\label{momentumconslaw}
\ee
and the variable $s$ can be approximated in the asymptotic limit by
$s\approx \mu_1^2 + \mu_2^2 + 2 (E_1 E_2 - {\bf P}_1 {\bf P}_2)$.

The simplest choice corresponds to $f=0$ leading to
conservation laws which take the standard SR additive form in terms of
the auxiliary variables. Then these laws can be trivially extended to a
system with an arbitrary number of particles.
This choice for $f$ could naively lead us to think
that we are dealing just with SR, only rewritten in some strange variables
(the auxiliary energy-momentum). This criticism has already been presented in
the DSR context (see, e.g., Ref.~\cite{jafari}), and answered several times
(see, e.g., Ref.~\cite{dsrfacts}). It is based on the wrong identification of
the existence of a mapping between the mathematical descriptions of two theories
with their
physical equivalence.\footnote{For instance, the existence of a canonical
transformation that relates the coordinates of the system of two coupled oscillators
to the coordinates of a system of two uncoupled oscillators obviously does not
establish their physical equivalence.} The main point is that the energy and
momentum
of a particle have an intrinsic meaning (for example, as the generators of
time and spatial translations) so that one is not allowed to define a `new'
energy from a nonlinear combination of the previous variables.

Another option however is to have $f\neq 0$. In this case it is not
guaranteed that the conservation law can be expressed as a sum of contributions
each one depending on the energy-momentum of only one particle. In this last instance,
Planckian effects make the two-particle system to
loose any kind of separability between their elements.\footnote{We
will say a conservation law is separable if
and only if there is an equivalent equation in which each member can be written as a
sum of terms each one depending just
on the energy-momentum of one particle. In the case of $f(x,y)=0$ one can reexpress
the
conservation
law as an equivalent equation in which there are terms which contain energy-momentum
of several particles (it is in this way as the conservation law is often written in
the DSR literature). However, it can be written in an equivalent way as a sum of
contributions each one depending on only one particle and, therefore, when $f(x,y)=0$
the conservation law is separable.} Then adding one more particle introduces
new
arbitrariness and the conservation laws have to be studied separately
for a given number of particles. In particular the associative
property is lost; it has been argued~\cite{jv}
that such kinematic generalizations should be ruled out but we do not
find any reason to exclude them within the rules used to define an
asymptotic approach to SR kinematics.

\subsubsection*{Domain of validity of ASR in the two-particle system}

The possibility to have generalized conservation laws may give rise to the peculiar
situation in which ASR is identical to Special Relativity at the one-particle
level (with unmodified dispersion relation and Lorentz transformations) but
different from it if one considers a two-particle system. This happens if
Planckian corrections are such that $g=h=0$ and $f\neq 0$. In this case the
ultra-small factor $s^{\alpha/2}/M_P^{\alpha}$ of Eq.~(\ref{auxenergyconslaw})
makes these corrections unobservable unless $f(\mu_1^2/s,\mu_2^2/s)$ takes
values much larger than one. To see how this may occur, first note that,
for two given particles, $\mu_1$,
$\mu_2$ fixed, $f$ is a function of only one variable, the Lorentz invariant
Mandelstam variable $s$. An amplification of $f$ compatible with experimental
observations may then only arise for values of $s$ outside of the range where
Special Relativity is well tested.
In the center of mass reference frame, $s$ is determined by the momentum $P$
of each of the two particles. The study of two-particle scattering has
experimental limitations both at arbitrary large values of $P$, what we call
the `high-energy' (ultraviolet, UV) regime, and at arbitrary low values of $P$
(since one cannot determine that the two particles are at rest with arbitrary
precision), what we refer to as the `low-energy' (infrared, IR) limit.
This means that we may have an amplification effect only in these two limits:
very large $s$ (but still $s\ll M_P$ in the framework of ASR), or very low $s$.
In the first case, one gets an amplification effect if $f$ is such that
\be
f\left(\frac{\mu_1^2}{s}, \frac{\mu_2^2}{s}\right)\to
\left(\frac{s}{\mu^2}\right)^{\gamma'},\ \gamma'>0, \quad \text{when }
s\gg \mu_1^2, \mu_2^2\,,
\label{UVamplif}
\ee
where $\mu^2$ is some combination of $\mu_1^2$ and $\mu_2^2$. Note that the UV regime
does not include, for example, the collision process between ultra-high energy cosmic
rays (of energy $E\sim 10^{21}\,$eV) and cosmic microwave background photons
(of energy $\omega\sim 10^{-3}\,$eV), since for this process $s$ is not
much larger than $\mu_N^2$ (the squared of the proton mass):
\be
s\approx \mu_N^2 + 2E\omega (1-\cos\theta)\quad \Rightarrow \quad
\frac{s}{\mu_N^2}\approx
1+2\,\frac{E}{10^{21}\,\text{eV}}\frac{\omega}{10^{-3}\,\text{eV}}
(1-\cos\theta)\,,
\ee
where $\theta$ is the angle between the directions of the momenta of proton and photon
in the laboratory reference frame.

The other scenario in which amplification effects for $f$ compatible with the
observed phenomenology may arise is in the IR limit. In the case of
two massive particles,
when $s$ approaches its lower bound,
$s\sim (\mu_1 + \mu_2)^2$, the arguments of $f$ take the values
$(\mu_1^2/(\mu_1+\mu_2)^2,\mu_2^2/(\mu_1+\mu_2)^2)$.
For a combination of two particles with given masses one could make
an \textit{ad hoc} choice for the function $f$ such that one had an amplification
effect that would be specific to this combination of particles. We will
consider the simpler and, as we will see, phenomenologically interesting case,
in which one of the particles is massive
with mass $\mu_1$ and the other is a zero mass particle, such as a photon.
The IR limit in this case corresponds to values of the arguments of
$f$ equal to 1 and 0,
independently of the value of $\mu_1$, and an amplification in the
infrared is obtained if
\be
f(x,0)\to (1-x)^{-\beta'},\ \beta'>0, \quad \text{when } x\to 1.
\label{IRamplif}
\ee

Consistency within the ASR framework requires that, in the case of an ultraviolet
amplification taking place at $s\gg \mu_1^2, \mu_2^2$ as in Eq.~(\ref{UVamplif}), the
modification in the conservation laws has to be small,
$(s^{\alpha/2}/M_P^\alpha)(s/\mu^2)^{\gamma'}\ll 1$
or $s\ll \mu^{2/(1+\alpha/2\gamma')}M_P^{2/(1+2\gamma'/\alpha)}$. This
defines a domain of validity of ASR in the
ultraviolet for a system of two particles which is smaller than the
naive domain $s\ll M_P^2$.

In the case of the infrared limit of a massive particle $\mu_1$ and a photon,
and with the infrared amplification Eq.~(\ref{IRamplif}),
ASR is limited to the domain where
$(s^{\alpha/2}/M_P^\alpha)(1-\mu_1^2/s)^{-\beta'}\ll 1$, or
$s-\mu_1^2\gg \mu_1^2(\mu_1/M_P)^{\alpha/\beta'}$.

\section{Some phenomenological implications of corrections to SR kinematics}

In order to illustrate the sensitivity of different observations to
departures from SR we are going to consider the spectrum of ultra high
energy cosmic rays (UHECR) as the best possibility to observe a departure from
SR with an amplification mechanism in the UV/ultrarelativistic limit, the frequency
shift in a double Raman transition, which is the most sensitive
observation to an amplification in the IR/nonrelativistic limit of a departure from SR,
and the effect of Planckian kinematic corrections in the energy
dependence of the velocity of propagation of a particle. In all these
cases we will take for definiteness a value $\alpha=1$ for the
exponent characterizing the asymptotic approach to SR so that the
Planckian corrections will be proportional to $(1/M_P)$.

\subsection{GZK cutoff}
We first consider a  separable conservation law with $f=0$. In this case
the kinematic analysis of SR applies directly to the auxiliary
variables. Let us remind the derivation of the threshold for the
energy loss of protons propagating in the cosmic microwave background
(CMB) through pion production. The dispersion relation of the proton,
in terms of auxiliary variables, is
\be
\epsilon \approx \pi + \frac{\mu_N^2}{2\pi} {\hskip 1cm} \pibf = \pi
{\bf n}\,,
\ee
with ${\bf n}$ a unit vector in the direction of propagation of the
proton. The threshold can be determined by considering in the final
state both the nucleon and the pion momenta in the same direction as
the initial proton so that
\be
\epsilon_1' \approx x_1 \pi + \frac{\mu_N^2}{2 x_1 \pi} {\hskip 1cm}
\pibf_1' \approx x_1 \pi {\bf n}
\ee
for the nucleon and
\be
\epsilon_2' \approx x_2 \pi + \frac{\mu_{\pi}^2}{2 x_2 \pi} {\hskip 1cm}
\pibf_2' \approx x_2 \pi {\bf n}
\ee
for the pion. Then energy conservation leads to
\be
\pi + \frac{\mu_N^2}{2\pi} + \omega \approx x_1 \pi +
\frac{\mu_N^2}{2 x_1 \pi} + x_2 \pi + \frac{\mu_{\pi}^2}{2 x_2 \pi}
\label{energycons}
\ee
and momentum conservation gives
\be
\pi - \omega = x_1 \pi + x_2 \pi\,,
\label{momentumcons}
\ee
where $\omega\sim 10^{-3}\,$eV is the energy of the CMB-photon whose momentum is
chosen in the opposite direction to the proton in order to determine
the threshold of the reaction. Terms proportional to the proton
momentum $\pi$ cancel when we subtract Eqs.~(\ref{energycons}) and (\ref{momentumcons}),
\be
\frac{\mu_N^2}{2\pi} + 2 \omega \approx  \frac{\mu_N^2}{2 x \pi} +
\frac{\mu_{\pi}^2}{2 (1-x) \pi}\,.
\ee
In the last step we have made the approximation $\pi - \omega \approx
\pi$ in the momentum conservation equation which then leads to $x_1 +
x_2 \approx 1$ and we have used the notation $x_1=x$. Then one has
\be
\pi = \frac{1}{4\omega} \left[\frac{\mu_N^2}{x} +
  \frac{\mu_{\pi}^2}{1-x} - \mu_N^2\right].
\ee
The threshold in the auxiliary variable
\be
\pi \geq \pi_{th} = \frac{\mu_{\pi} (\mu_{\pi}+2\mu_N)}{4 \omega}
\label{GZKthreshold}
\ee
coincides with the SR-threshold ($P_{th}^{SR}$). The Planckian
correction to the momentum threshold is
\be
\frac{P_{th}}{P_{th}^{SR}} - 1 \,=\, \frac{\epsilon_{th}}{M_P} \,
h\left(\frac{\pi_{th}}{\epsilon_{th}}\right)\,.
\label{GZKcorrection}
\ee
It is clear that
in order to have an appreciable effect of departures from SR in the
cosmic ray (CR) spectrum at very high energies it is necessary to
consider a generalized kinematics with $|h(x)|\gg 1$ when $x\approx
1$. As in Eq.~(\ref{URlimit}) we can consider
\be
h(x) \approx \eta_{\gamma} (1 - x)^{-\gamma} \,\,\, \text{when}
\,\,\,  x \approx 1\,,
\label{GZKURlimit}
\ee
with $\gamma$ a positive real number and $\eta_{\gamma}$ a constant
coefficient which can be used as a measure of the energy scale
characterizing the kinematic corrections. A value of
$\eta_{\gamma}$ of order one means an energy scale of the order of
magnitude of the Planck mass as expected for an effect due to quantum gravity.
Inserting Eq.~(\ref{GZKURlimit}) into Eq.~(\ref{GZKcorrection}) we get
\be
\frac{P_{th}}{P_{th}^{SR}} - 1 \,\approx\,
\eta_\gamma\, 2^\gamma
\left(\frac{\mu_N}{M_P}\right)
\left[\frac{\mu_{\pi}}{2\omega}
  \left(1+\frac{\mu_{\pi}}{2\mu_N}\right)\right]^{(2\gamma+1)}\,,
\ee
and, since the quantity between square brackets is of the order $10^{11}$, there
may have observable corrections for $\eta_\gamma$ of order one (i.e. Planckian effects)
and even
much smaller values depending on the value of the exponent $\gamma$ which controls
the amplification of the Planckian corrections. One can arrive to the same conclusions
directly from the expression of the momentum in terms of auxiliary
variables
\begin{equation}
P \,=\, \pi \left[1 + \eta_{\gamma} \frac{\epsilon}{M_{P}}
\left(1-\frac{\pi}{\epsilon}\right)^{-\gamma}\right] \approx
\pi \left[1 + 2^{\gamma}\, \eta_{\gamma} \, \frac{\mu_N}{M_{P}}
\left(\frac{E}{\mu_N}\right)^{2\gamma+1}\right]
\end{equation}
where in the second step we have assumed $\epsilon \gg \mu_N$ and we
have made the approximation $\epsilon \approx E$ in the Planckian
correction. In this way one can see that varying $\gamma$ one changes
the domain of energies where the difference between the momentum
variable and the corresponding auxiliary variable is appreciable
and then one can have signals of departures from SR kinematics in the
UHECR spectrum.

As an alternative to the previous example where one gets an observable
effect in the cosmic ray spectrum as a consequence of an appreciable
difference between the momentum variable and the auxiliary variable
$\pi$ at scales much lower than the Planck mass, let us consider an
example were $g=h=0$ and then there is no correction in the kinematic
description of a one particle system. In this case all the departures
from SR are characterized by a function $f$.
But as it has already been pointed
out in the previous Section, the natural suppression of Planckian
kinematic corrections by a factor ${\sqrt s}/M_{P}$ and the
cancellations due to the relativity principle can not be
compensated by an amplification in the ultraviolet limit, something which would
require to go to energies far beyond those explored in CR physics. An
amplification in the infrared limit would require to
consider energies far below the energy range where the CMB background
has any appreciable effect in the CR spectrum. To summarize the UHECR
spectrum is not affected by this type of departures from SR.

\subsubsection*{Some comments on ASR compared to DSR}

Besides the obvious suppression in kinematic Planckian corrections due to the
smallness of the highest accessible energy in microscopic systems with respect to
the Planck scale (the GZK threshold is nine orders of magnitude lower than the
Planck energy), in theories that preserve a relativity principle
there is another source of suppression in the form of cancellations of Planckian effects
owing to the consistency between the conservation laws of the energy-momentum,
the dispersion relation
and the relativity principle. For this reason, in general, one has no
signals on the UHECR spectrum
of a departure from SR compatible with a relativity principle unless
one makes special choices
of the deformed kinematics. One explicit example
of this situation corresponds to the family of DSR proposals
which produce an observable change in the prediction of the GZK
cut-off (see Ref.~\cite{dsr3}). One can ask which is the relation
between these proposals
and the UV amplification mechanism introduced in the ASR framework. As
it has already been said
in the Introduction, the ASR idea is closely
related to the leading contribution of DSR at scales much lower than the Planck scale.
In fact, the leading terms of all DSR proposals compatible with the
treatment of conservation laws in Ref.~\cite{jv} are particular
cases of ASR with $f=0$.\footnote{Note that, according to its definition,
ASR may include non-analytical behaviour in the modified dispersion
relation. In this aspect it also goes beyond the DSR theories
constructed so far.}
In the case of the DSR proposals of Ref.~\cite{dsr3},
their leading term reduces to the form of Eq.~(\ref{GZKURlimit}) at
low enough energies.
Of course, different DSR models can have the same behavior at low energies
(that is to say, their leading-order correction to SR can be the
same) and therefore they converge to the same ASR.
The reason why Planckian effects become
relevant in these proposals is related to the amplification
mechanisms discussed in this work.

With respect to our discussion of modified conservation laws,
we have not found any  DSR proposal corresponding to $f\neq 0$.
We think that the analysis presented in this work could be of help
to introduce such a generalization in the DSR framework.
One should note however that our analysis excludes cases like conservation
laws where the dependence on initial and final state variables cannot be separated
as it happens in certain DSR proposals based on attemps to construct a relativistic
theory in $\kappa$-Minkowski.

\subsection{Atomic interferometry}

We now consider the experiment with greater sensitivity to kinematic
corrections in the nonrelativistic/IR limit. The possibility to identify
a deviation from SR predictions in atomic interferometry has received
some attention recently. A first analysis of the
sensitivity of these experiments within a simple parametrization of a
generalized energy-momentum relation for an atom in the
nonrelativistic limit has lead~\cite{mercati} to the remarkable
conclusion that one could detect
Planckian kinematic corrections thanks to the extreme precision of the
determination of the
difference between the absorbed and emitted frequencies by cold atoms and
the possibility to observe these transitions with very slow atoms. It
is then natural to discuss this problem in the framework introduced in
this work.

The process to consider is the absorption of a photon by an atom, both
moving in the same direction, and posterior emission of a photon in the
opposite direction. In the case of a  separable generalized
conservation law ($f=0$) once more one can repeat the standard
kinematic analysis of this process using the auxiliary variables. The
energy conservation law gives
\be
M + \frac{\pi^2}{2 M} + \nu \approx M + \frac{\pi'^2}{2 M} + \nu'
\ee
and for the momentum conservation law one has
\be
\pi + \nu \approx \pi' - \nu'\,,
\ee
where $M$ is the mass of the atom, $\nu$ ($\nu'$) the frequency of
the absorbed (emitted) photon and $\pi$ ($\pi'$) the auxiliary
variable of the initial (final) atom. We have neglected the subleading
effect due to kinematic corrections to the description of the absorbed
and emitted photons (see Ref. \cite{mercati}). One can also
approximate at the level of the
momentum conservation law $\nu \approx \nu' \approx \nu^{*}$, with
$\nu^{*}$ the Bohr frequency associated with the intermediate excited
atom. Then from the energy conservation law one has
\be
\frac{\pi^2}{2 M} + \nu \approx \frac{(\pi + 2 \nu^{*})^2}{2 M} +
\nu'
\ee
and the frequency shift is given by
\be
\Delta \nu = \nu - \nu' = \frac{2\nu^{*} (\nu^{*} + \pi)}{M}\,.
\ee
We can reexpress the frequency shift as a function of the momentum $P$
of the atom using the nonrelativistic approximation $E\approx M$
in the terms suppressed by the Planck mass of Eq.~(\ref{aav}):
\be
\Delta \nu = \Delta\nu_{SR} - \frac{\nu^{*} P}{M_P} h(P/M)\,,
\ee
with $\Delta\nu_{SR} = 2\nu^{*}(\nu^{*} + P)/M$. The
correction to the frequency shift is
\be
\frac{\Delta \nu}{\Delta\nu_{SR}} - 1 \,=\, -\frac{M}{2 M_P}
(1+\nu^{*}/P)^{-1}\, h(P/M)\,.
\label{f}
\ee
The suppression factor $M/M_P$ makes this kinematic
correction totally unobservable unless one has $h(x)\gg 1$ when
$x\ll 1$, partially compensating the Planck mass suppression and
leading to a correction of the order of magnitude of the precision in
the determination of the frequency shift. A simple example with an
amplification of the kinematic correction in the nonrelativistic limit is
that of Eq.~(\ref{NRlimit}),
\be
h(x) \approx \xi_{\beta}\, x^{-\beta} \,\, \text{when} \,\, x\ll 1\,,
\ee
leading to
\be \label{deltanubeta}
\frac{\Delta \nu}{\Delta\nu_{SR}} - 1 \,\approx\, - \xi_{\beta} \frac{M}{2
  M_P} (1+\nu^{*}/P)^{-1} \frac{M^{\beta}}{P^{\beta}}\,.
\ee
Introducing a value $10^{-8}$ for the correction to the frequency
shift (\ref{f}) as an observability criteria we find that the
corresponding value for $\xi_{\beta}$ is
\be
\xi_{\beta} \approx \pm 10^{-8}\, \frac{2M_P}{M} (1+\nu^{*}/P)
\frac{P^{\beta}}{M^{\beta}}\,,
\ee
which is of order one (Planckian effect) for $\beta=1$ and
$P\sim \text{100 eV}$.
In this way one
reproduces, within the present general kinematic framework compatible
with the relativity principle, the result that atomic interferometry
experiments can be sensitive to kinematic corrections with an energy
scale of the order of the Planck mass~\cite{mercati}.

In analogy with the discussion of kinematic corrections in the
UV limit, one can consider the possibility to have an
appreciable effect in the IR limit in a situation where
there is no modification of the energy-momentum relation for the atom
and the only kinematic correction appears at the level of the
generalized conservation laws.
In this case one has
\be
\frac{P^2}{2 M} + \nu + \frac{{\sqrt s}}{M_{P}} M f(M^2/s, 0)
\,\approx\,  \frac{(P+2\nu^{*})^2}{2 M} +
\nu' + \frac{\sqrt{s'}}{M_{P}} M f(M^2/s', 0)\,,
\label{deltanuf}
\ee
which is the generalized energy conservation law Eq.~(\ref{energyconslaw})
particularized to the case $\alpha =1$ (Planckian corrections
proportional to $1/M_{P}$) and $\mu_2=0$ (one massless particle), and where
we have neglected $E_2 f$ in comparison with $E_1 f\approx Mf$. We
also have
\be
s \approx M^2 + 2 M \nu \,, \quad \quad s' \approx M^2 + 2 M \nu'\, .
\ee
It is remarkable that now there is a momentum (of the atom)
independent contribution in the energy conservation equation which
does not cancel. This is the main
reason why one can easily find choices of the modified conservation
laws (function $f$) leading to observable effects in the frequency
shift.

As a first example we consider a constant function $f = \rho_0$. In this
case one finds from Eq.~(\ref{deltanuf}) that
\be
\Delta\nu \approx \Delta\nu_{SR} \left[1 - \rho_0 \frac{M}{M_{P}}\right]
\ee
and then Planckian corrections are unobservable unless $\rho_0 > 10^9$ (we
are assuming a precision on the determination of $\Delta\nu$ of order
$10^{-8}$). Next we can consider $f(x,0)=\rho_{\beta'} (1-x)^{-\beta'}$
[see Eq.~(\ref{IRamplif})]
which gives an amplification of the Planckian corrections
\be
\Delta\nu \approx \Delta\nu_{SR} \left[1 + \frac{\beta'
    \rho_{\beta'}}{2^{\beta'}} \left(\frac{M}{\nu^*}\right)^{1+\beta'}
  \frac{M}{M_{P}}\right]
\ee
such that, for typical values $M/\nu^*\sim 10^{11}$, one can have an observable
effect of Planckian corrections
for $\rho_{\beta'}$ of order one (Planckian sensitivity) and even much
smaller values depending
on the choice of the exponent $\beta'$ which controls the
amplification of the Planckian corrections.

\subsection{Velocities}
To end up this exploratory analysis of possible ways to detect
implications of ASR we can consider observations which are sensitive to
corrections in the energy dependence of the velocity of
propagation. In fact, the strongest constraints to departures from SR
come in many cases~\cite{v(E)} from limits to an energy dependence for
the light propagation velocity and its consequences in cosmic gamma
rays. In order to discuss these limits it is
required to go beyond the framework we have considered until now.
The previous analysis was based exclusively on the kinematic
description of particles with energy-momentum variables and the
relations among these variables (conservation laws) for processes with
initial and final states with up to two particles. The simplest way to
extend the discussion considering the propagation of a particle in
space-time is to add the assumption\footnote{In fact the structure of
space-time in different attempts to go beyond SR is an open issue.}
that the energy-momentum variables used in the kinematic description
of the one particle systems are the generators of translations in
space-time. In this case one has
\be
\frac{d\xbf}{dt} \,=\, \frac{\partial E}{\partial \pbf} \,=\,
\frac{\pbf}{E} \left[1 + \frac{3E}{2M_{P}} \Delta(p/E) +
  \frac{E}{2M_{P}} \left(\frac{E}{p}-\frac{p}{E}\right)
    \Delta'(p/E)\right]\,,
\ee
where we have used the asymptotic dispersion relation (\ref{adr}) with
$\alpha=1$. If one considers an amplification in the nonrelativistic
and/or ultrarelativistic limit in order to have some appreciable effect
of the kinematic Planckian corrections, then one can use the
corresponding choices for the functions $g$, $h$ to calculate the
function $\Delta$ which fixes the modification in the dispersion
relation and the velocity of propagation of particles. A systematic
analysis of the modification of velocities in different limits and the
possibility to observe them will be presented elsewhere.

In the case of photons, the dispersion relation (\ref{adr0}) leads to
\be
c(E) \,=\, \left|\frac{d\xbf}{dt}\right| \,=\, 1 + \frac{E}{M_{P}}
  ({\tilde g}- {\tilde h})\,,
\ee
which is the standard energy dependence for the velocity of light as
expected on pure dimensional grounds. From different observations of
cosmic gamma rays one has very stringent limits on $(\tilde{g}-
\tilde{h})$. It should be noted that these limits do not have any
implication on the previous discussion of signals of departures from
SR in the UHECR spectrum and atomic interferometry because the
modifications on the photon kinematics were irrelevant in those
cases.

\section{Concluding remarks}

We have introduced a convenient parametrization for the asymptotic
approach to SR kinematics compatible with the relativity principle. In
the simplest case of a separable generalized energy-momentum
conservation law it is given in terms of two functions of one variable
which fix at the same time the modification of the dispersion relation
and the conservation laws. It is also possible to have nonseparable
contributions in the generalized conservation laws; these contributions
require then the introduction of an additional function of two
variables. In fact it is possible to have a generalized conservation
law with the standard SR dispersion relation (setting the two
functions of one variable equal to zero) while maintaining the
relativity principle. This provides a general framework
within which to explore possible signs of
the space-time structure of a theory of quantum gravity.

We have shown how it is possible to escape from the difficulty to observe
these signs at energies much smaller than the Planck scale.
The cancellation of corrections in an approach compatible with the
relativity principle can be (at least partially) compensated by an
amplification mechanism leading to observable effects at energies much
lower than the Planck mass. Two limits where one can introduce an
amplification are the UV/IR limits (particles with momentum much
higher/lower than the mass). An amplification in the UV limit implies
a reduction in the maximum energy-momentum of the domain of validity
of the asymptotic kinematic description and an amplification in the IR
limit leads to the striking conclusion that the Planckian corrected
kinematics cannot be applied to arbitrarily small momenta.

We have applied the proposal for a generalization of SR kinematics to
the simplest reaction responsible for the energy loss of protons
propagating in the CMB. We have  found simple examples of an
amplification in the UV limit leading to observable effects in the
spectrum of very high energy cosmic rays due to the modification of
the dispersion relation for protons. The generalized kinematics has
also been used to calculate the frequency shift in a double-Raman
atomic transition. In this case we have identified simple examples of
an amplification in the IR limit with appreciable corrections to the
SR result due either to a modification in the momentum dependence of
the kinetic energy of the atom or to a non-separable generalization of
the energy-momentum conservation law for a photon-atom system.
Once the very simple analysis based on the results for the momentum
threshold of the photopion production from protons and the frequency
shift in a double Raman transition has lead us to the conclusion that
there can be appreciable effects of Planckian kinematic corrections
both at high and low momenta one can go further to try to extract
information of the details of these corrections from present and/or
future data. This requires to go beyond the simple analysis
presented in this work.

In this direction, at high momenta one should try to extend the
usual analysis of the high energy cosmic ray spectrum based on SR
kinematics incorporating the Planckian corrections. It would also be
interesting to explore other reactions where the amplification
introduced in the UV limit could lead to appreciable effects; one case
which seems specially promising is the scattering of very high energy (TeV)
neutrinos where the very small value of the neutrino masses makes the
amplification more important.

At low momenta it could be interesting to extend the analysis of cold
atom experiments exploring the possibility to get complementary
information on the Planckian corrections by considering different
setups corresponding to different directions for the absorbed and
emitted photons. The analysis of properties in particle reactions
which are sensitive to IR divergences is another way to look for data
which can be sensitive to kinematic Planckian corrections. Another
possibility to go towards the limits of validity of the generalized
kinematic framework and then to the onset of appreciable corrections
is to consider systems at extremely low temperatures like
Bose-Einstein condensates.
When comparing the different strategies to look for signs of kinematic
Planckian corrections one should keep in mind that the most
straightforward way, the approach to the limits of validity of the
simple asymptotic kinematic description, is limited by the increase
of the unknown corrections to the asymptotic limit. Alternatively, high
precision data can be sensitive to a small correction well within the
domain of validity of the asymptotic description and then such
kind of experiments are better
candidates to determine the details of the asymptotic approach to
SR. Of course the primordial objective is the identification of a
departure from SR and the best strategy to get this result depends on
the amplification mechanisms accompanying these departures.

\section*{Acknowledgments}
We would like to thank A. Grillo for discussions. This work is
supported by CICYT (grant FPA2009-09638) and DGIID-DGA (grant
2009-E24/2). D.M. acknowledges a FPI grant from MICINN.



\end{document}